\documentclass[10pt,preprint]{aastex}

\newcommand{\be}{\begin{equation}}
\newcommand{\ee}{\end{equation}}
\newcommand{\ba}{\begin{eqnarray}}
\newcommand{\ea}{\end{eqnarray}}
\newcommand{\siml}{\lower4pt \hbox{$\buildrel < \over \sim$}}
\newcommand{\simg}{\lower4pt \hbox{$\buildrel > \over \sim$}}

\begin{document}

\title{Reply to Lyutikov's comments on Zhang \& Kobayashi (2005)}

\author{Bing Zhang\altaffilmark{1} and Shiho Kobayashi\altaffilmark{2,3}}
\affil{
$^{1}$Dept. of Physics, University of Nevada, Las Vegas, NV 89154
\\
$^{2}$Dept. of Astronomy \& Astrophysics, Penn State
University, University Park, PA 16802 \\
$^{3}$Dept. of Physics, Penn State University,
University Park, PA 16802}

\begin{abstract}

Lyutikov (astro-ph/0503505) raised a valid point that for shock
deceleration of a highly magnetized outflow, the fate of the magnetic
fields after shock crossing should be considered. However, his comment
that the deceleration radius should be defined by the total energy
rather than by the baryonic kinetic energy is incorrect. As strictly
derived from the shock jump conditions in Zhang \& Kobayashi (2005),
during the reverse shock crossing process the magnetic energy is not
tapped. As a result, the fireball deceleration radius is defined by
the baryonic energy only. The magnetic energy is expected to be
transferred to the circumburst medium after the reverse shock
disappears. The evolution of the system then mimicks a
continuously-fed fireball. As a result, Lyutikov's naive conclusion
that the forward shock dynamics is independent on the ejecta content
is also incorrect. The shock deceleration dynamics and the reverse
shock calculation presented in Zhang \& Kobayashi (2005) are robust
and correct. 

\end{abstract}

In our recent paper Zhang \& Kobayashi (2005, hereafter ZK05), we
calculated the gamma-ray burst (GRB) reverse shock emission when a
highly relativistic outflow with a wide range of magnetization
parameter $\sigma$ interacts with a constant density medium
(ISM). We found that throughout the period when the reverse shock
crosses the ejecta, the lab-frame Poynting flux energy is essentially
unchanged. Considering a fireball with a total energy
$E=E_{K,0}+E_{P,0} = E_{K,0} (1+\sigma)$, where $E_{K,0} =\gamma_0 M_0
c^2$ is the initial baryonic kinetic energy in the ejecta and
$E_{P,0}$ is the initial Poynting flux energy, we drew the conclusion
that  the radius at which the fireball collects $1/\Gamma_0$ of the
initial fireball rest mass, i.e.
\be
R_\gamma \sim R_{dec} \sim \left(\frac{E_{K,0}}{\gamma_0^2 n m_p
c^2}\right)^{1/3}~, 
\label{Rgam}
\ee 
defines the radius for fireball deceleration in the ``thin shell''
regime (Eq.[41] in ZK05) when a reverse shock exists. This is smaller
than the conventional deceleration radius (defined by $E$, in the
$\sigma=0$ limit) by a factor $(1+\sigma)^{1/3}$. As a result, the
forward shock bolometric emission level right after this deceleration
radius is correspondingly fainter by a factor of $(1+\sigma)$
comparing to $\sigma=0$ case. 

In a recent astro-ph comment by Lyutikov (2005), he raised a valid
point that the fate of magnetic fields after shock crossing, which has
been ignored in the version 2 of our paper posted in astro-ph, should
be considered. The Poynting energy ($E_{P,0}$) should be eventually
transferred to the ISM. We have incorporated this
comment in our final version to appear in ApJ (version 3 in astro-ph),
and discussed the possibility of transferring $E_{P,0}$ to the ISM after
the shock crossing stage. 

However, there exist severe errors in Lyutikov's comments, which would
mislead the readers if not corrected. 

The major error in Lyutikov (2005) is that he defines deceleration
using the total energy $E$ regardless of the $\sigma$ value. This
gives
\be
R_{dec}^{L} \sim \left(\frac{E}{\gamma_0^2 n m_p
c^2}\right)^{1/3} = R_{dec} (1+\sigma)^{1/3}~.
\label{RdecL}
\ee 
Such a definition is {\em purely based on his intuition without
investigating the rigorous shock jump conditions carefully.}  The
consequence of adopting such {\em an assumption} unfortunately leads
to self-contradictory results in the high-$\sigma$ regime. We now
elaborate it in the following.

There are several characteristic radii when discussing the
deceleration of a fireball by ISM through reverse shocks (Sari \&
Piran 1995). Besides $R_\gamma$, the relevant ones also include
$R_\Delta$, the radius at which the reverse shock crosses the ejecta
shell, and $R_N$, the radius at which the reverse shock upstream and
downstream becomes relativistic to each other. When $\sigma$
increases, a reverse shock forms when the forward shock pressure
exceeds the magnetic pressure in the outflow (Eqs.[31] and [43] in
ZK05). The critical radii discussed above may deviate from the values
in the $\sigma=0$ limit. A detailed analysis shows that for a fixed
total energy $E$, $R_N$ does not depend on $\sigma$, and $R_\Delta$ is
roughly smaller by a factor $\sigma^{1/2}$ (Eqs [36] and [38] in
ZK05). The relations among these critical radii are characterized by
Eqs. (46) or (51) in ZK05. One can then categorize various parameter
regimes in the $T/t_\gamma - \sigma$ plane as long as the reverse
shock forming condition is satisfied (see Fig.4 of ZK05), where $T$ is
the duration of the burst, $t_\gamma$ is a typical time scale defined
by the total energy $E$ and the initial Lorentz factor $\gamma_0$
(Eq.[48] in ZK05, where $\gamma_0$ was denoted as $\gamma_4$).

Regardless of the $\sigma$ value, as long as $T$ is long enough
(Region (I) in ZK05), one should be in the regime of $R_N < R_\gamma <
R_\Delta$, i.e. the so-called thick shell regime (cf. Sari \& Piran
1995 for the case in the $\sigma=0$ limit). In this regime, the
reverse shock upstream and downstream becomes relativistic with each
other before the reverse shock crosses the shell. As a result, the
ejecta is significanly decelerated after one shock crossing. In this
regime $R_\Delta$ defines the radius of fireball
deceleration. $R_\gamma$ is no longer meaningful.

As $T$ becomes smaller, one enters the thin shell regime. In
this regime, shock crossing only mildly decelerates the shell, and the
shell starts to decelerate significantly when it feels a large enough
inertia from the ISM (at $R_{dec}$). In this regime $R_{dec} \leq R_N$
should be satisfied. In the $\sigma=0$ limit, this
naturally happens when $T < t_\gamma$. When a higher $\sigma$
value is considered, one still gets a consistent picture when
$R_{dec}$ is defined as $R_\gamma$ 
(Eq.[\ref{Rgam}]), and the transition naturally happens at $T <
t_\gamma Q(\sigma)$, where $Q(\sigma)$ is a function of $\sigma$
($\propto \sigma^{2/3}$ at large $\sigma$). However, if one uses the
un-justified deceleration radius defined by Lyutikov, $R_{dec}^L$
(Eq.[\ref{RdecL}]), one reaches an unphysical regime
$R_N < R_\Delta < R_{dec}^L$ when $\sigma^{2/3} \leq T/t_\gamma \leq
\sigma^2$ is satisified. In such a regime, the reverse shock 
upstream and downstream becomes relativistic first (at $R_N$), so that
after the reverse shock crosses the shell (at $R_\Delta$), the shell
should be significantly decelerated. However, the fireball has not
reached $R_{dec}^L$ yet - the radius Lyutikov expects the fireball to
decelerate. This conclusion is obviously absurd, which already
invalids his definition. Below we will explicitly analyze the
inconsistency in Lyutikov's argument.

Defining $R_\gamma$ as $R_{dec}$ in the thin shell regime by ZK05 {\em
is not an assumption}. It stems from solid energy conservation
physics and rigorous shock jump conditions. Let us analyze the
deceleration process in detail. 

(1) Before the shock crossing process is over, the system contains four
regions separated by two shocks and one contact continuity. One can
write the energy conservation during the process. Assuming that the
ISM mass collected by the fireball at the end of shock crossing is
$M_{\rm ISM}$, The total energy of the whole system before the shock
crossing is $\gamma_0 M_0 c^2 + E_{P,0} + M_{\rm ISM}c^2$. After shock
crossing, the total energy of the system is $\gamma (\gamma_{34} M_0 +
\gamma M_{\rm ISM}) c^2 + E_P$, where $\gamma$ is the bulk Lorentz
factor at the end of shock crossing, $\gamma_{34}$ is the reverse
shock Lorentz factor which is $\sim 1$ in the thin shell case we are
discussing, the term $\gamma M_{\rm ISM}c^2$ takes into account the
rest ISM mass and its thermal energy (which is $(\gamma-1) M_{\rm ISM}
c^2$), and $E_P$ is the lab-frame Poynting energy after the shock
crossing. The energy conservation for the whole system then reads 
\be
(\gamma_0 - \gamma) M_0 c^2 + (E_{P,0}-E_P) = (\gamma^2-1) M_{\rm ISM}
c^2 ~.
\label{energy}
\ee
According to Eq.(40) of ZK05, strict shock jump condition
analysis gives $E_{P,0} \simeq E_P$ in the high-$\sigma$ regime. The
physical reason is that during the shock crossing the magnetic fields
are compressed so that their comoving energy increases. In the
meantime, the whole system is decelerated, so that the
lab frame Poynting energy $E_P = \gamma U'_B$ (where $U'_B$ is the
comoving magnetic energy) remains essentially unchanged. This means
that the term $(E_{P,0}-E_P)$ drops out from the Eq.(\ref{energy}).
The remaining equation is the standard fireball deceleration equation
with an effective fireball energy $E_{K,0}$. The fireball starts to
decelerate when $M_{\rm ISM} \sim M_0 / \gamma_0$ is satisfied, at the
radius defined by $R_\gamma$ (eq.[\ref{Rgam}]), not at $R_{dec}^L$
defined by Lyutikov (eq.[\ref{RdecL}]). 

(2) What is the status of the ejecta beyond
$R_\gamma$? What happens after the reverse shock
disappears? Is $R_{dec}^L$ meaningful at all? 

Again one can write down the energy conservation of the whole system
during the process. Let's simply re-define $\gamma_0$, $M_0$,
$E_{P,0}$ as the bulk Lorentz factor, effective mass and Poynting
energy at $R_\gamma$ (Eq.[\ref{Rgam}]), and $\gamma$, $E_P$ are the
Lorentz factor and Poynting energy at a later radius $R$. 
We also re-define $M_{\rm ISM}$ as the ISM mass 
collected during the process (from $R_\gamma$ to $R$). Then
the energy conservation can be still expressed as equation
(\ref{energy}). We show below that $\gamma$ must be significantly
smaller than $\gamma_0$. In other words, the fireball is decelerated
beyond $R_\gamma$.

If multi-crossing of the reverse shock still happens
beyond $R_\gamma$, the $(E_{P,0}-E_P)$ term still drops out from the
energy conservation equation, so that the fireball must be
decelerating. The energy transfer process only happens when the
reverse shock disappears.

During the energy transfer epoch (after the reverse shock
disappears), fireball deceleration is still going on. This is
manifested by the argument below. First, it is obvious that one
should always have $(E_{P,0}-E_P) \geq 0$, i.e. the Poynting energy
should only decrease and gets transferred to the medium. Next, the lab
frame Poynting energy could be written as  
\be
E_P = \gamma U'_B = \gamma V' \frac{B'^2}{8 \pi} = \gamma \frac{R^3} 
{\gamma} \frac{B'^2}{8 \pi} = R^3  \frac{B'^2}{8 \pi} ~,
\label{EP}
\ee
where $V'$ is the comoving volumn, which can be expressed as
$R^3/\gamma$ in the spreading phase (which is the case in the Regime
III of Fig.4 of ZK05). For the non-spreading case, a similar conclusion
could be also drawn.

If the GRB outflow does not decelerate from $R_\gamma$ to $R_{dec}^L$,
as Lyutikov hopes, one would maintain a forward shock with a constant
thermal pressure in the shocked region. The magnetic pressure
$B'^2/8\pi$ behind the contact discontinuity should keep constant in
order to maintain hydrodynamical equilibrium. Eq.(\ref{EP}) then
leads to an absurd conclusion, i.e. the lab-frame Poynting energy
increases with time (by a huge factor). 
This again invalidates {\em the assumption
of Lyutikov} that the outflow decelerates at $R_{dec}^L$. 

Then how is the Poynting energy transferred to the ISM eventually?
This is an interesting problem and needs further detailed
investigation. The treatment in ZK05 (ignoring this effect) is a
reasonable approximation shortly after the reverse shock peak, but
will become less rigorous when the reverse shock disappears because of
the energy transfer from the Poynting 
energy to the ISM. According to the energy conservation equation 
(\ref{energy}), one can see that the dynamics depends on how the term
$(E_{P,0}-E_P)$ evolves with $\gamma$. 
In any case, the system would mimick a continuously-fed
fireball, with the Poynting energy gradually injected into the system.
Physically, $B'^2/8 \pi$ is related to the pressure at the contact
discontinuity during the deceleration phase. Right after the shock
crossing (at or beyond $R_\gamma$ for thin shells and at $R_\Delta$
for thick shells), the pressure is balanced at the
contact discontinuity, and there is no net energy transfer from the
Poynting energy to the ISM kinetic energy. As the system decelerates,
the magnetic pressure behind the contact discontinuity becomes
stronger than the thermal pressure in front,
so it does work to the shocked region in front of the contact
discontinuity. This makes the system decelerates less significantly
than it would have been without such a magnetic pushing. Such a
process keeps going on, 
until most of the Poynting energy is transferred to the
ISM. The late afterglow of a high-$\sigma$ flow could be comparable to
that of a low-$\sigma$ flow, but in the early afterglow phase, the
afterglow level is significantly lower. The crucial point here is that
energy transfer phase is separated from the reverse shock
deceleration phase, so that the bolometric reverse shock emission peak
epoch is separated from the bolometric forward shock emission peak
epoch. Detailed deceleration process is under investigation.
One thing is clear: nothing special happens at the radius $R_{dec}^L$. 

It is now evident that another statement in Lyutikov (2005),
i.e. ``energy in the forward shock is determined by the total energy
and is virtually independent of the content of the ejecta'', is also
incorrect. According to the above analysis, the higher the $\sigma$,
the more prominant the injection process would be. Different $\sigma$
leads to different deceleration dynamics and different afterglow
lightcurves. The ``universal'' dynamics $\gamma \propto R^{-3/2}$
suggested by Lyutikov (notice he used $\propto t^{-3/2}$ so that his
$t$ is not the observer's time) is only valid when the energy transfer
process is over, i.e. when essentially all the energy of the system is
given to the ISM. However, the energy transfer is a
complicated process, and it obviously varies when $\sigma$ varies. The
naive treatment of Lyutikov (2005, and his previous work) 
lacks solid physical justification. 

In conclusion, the dynamics of shock deceleration of a
magnetized, relativistic outflow presented in ZK05 is
robust. The calculations of the reverse shock emission (which is the
main subject of our paper as reflected from the title) are correct.
The forward shock emission after the reverse shock disappears is not
treated in detail, and will be studied more carefully in a future
work. We thank Lyutikov\footnote{There have been extensive email
discussions between Lyutikov and us about the subject. Later we
decided to temporarily 
withdraw the paper from ApJ to add in a discussion about the caveat of
the forward shock calculation in our paper, and we planned to update
the astro-ph version after the paper is finalized. We have also
acknowledged Lyutikov for his critical comments, and let him know that we
are preparing the final version and will send it to him for comments when it
is completed. It is to our surprise that Lyutikov still decided to
expose his incorrect opinion at astro-ph. Although astro-ph might be a
chat board to exchange ideas, we believe that refereed
journals are more appropriate places to lay out scientific view
points. We will not reply to any further comments on this subject in
astro-ph.} for pointing out the flaw in our previous version of
ignoring the fate of the magnetic fields 
after shock crossing, and we have added a relevant discussion in the
final version of the paper to appear in ApJ (version 3 in
astro-ph). Yet, the flaw does not influence the bulk of the rigorous
calculations presented in the paper. In contrast, the main arguments
presented in Lyutikov (2005) are incorrect, as has been explicitly
explained in this reply.

\end{document}